 \documentclass{emulateapj}
 \usepackage{apjfonts}
\newcommand{\xmm}{\textit{XMM-Newton}}
\newcommand{\chan}{\textit{Chandra}}
\newcommand{\scrb}{$\sigma^2$ CrB}

\newcommand{\cmcc}{cm$^{-3}$}
\newcommand{\cmsq}{cm$^{-2}$}
\newcommand{\ergps}{erg~s$^{-1}$}
\newcommand{\flx}{ph~cm$^{-2}$~s$^{-1}$}

\newcounter{subfigure}

\submitted{To appear in the Astrophysical Journal}

\shorttitle{An \xmm\  Study of $\sigma^2$ CrB}
\shortauthors{Suh et al.}

\begin{document}


\title{An \xmm\  Study of the Coronae of $\sigma^2$ Coronae Borealis}

\author{Jin A. Suh and Marc Audard\altaffilmark{1}}
\affil{Columbia Astrophysics Laboratory, Mail code 5247, 550 West 120th Street, New York, NY 10027}
\email{jas2010@columbia.edu, audard@astro.columbia.edu}
\altaffiltext{1}{contact author}

\and

\author{Manuel G\"udel}
\affil{Paul Scherrer Institut, 5232 Villigen PSI, Switzerland}
\email{guedel@astro.phys.ethz.ch}

\and

\author{Frederik B.~S. Paerels}
\affil{Columbia Astrophysics Laboratory, Mail code 5247, 550 West 120th Street, New York, NY 10027}
\email{frits@astro.columbia.edu}

\begin{abstract}
We present results of \xmm\  Guaranteed Time
observations of the RS CVn binary $\sigma^2$ Coronae Borealis. 
The spectra obtained with the Reflection
Grating Spectrometers and the European Photon Imaging
Camera MOS2 were simultaneously fitted with collisional
ionization equilibrium plasma models to determine coronal
abundances of various elements. Contrary to the solar first ionization
potential (FIP) effect in which elements with a low FIP
 are overabundant in the corona compared to
the solar photosphere, and contrary to the ``inverse'' FIP
effect observed in several active RS CVn binaries, coronal
abundance ratios in \scrb\  show a complex pattern as supported by similar
findings in the \textit{Chandra} HETGS analysis of \scrb\  with
a different methodology \citep{osten03}. Low-FIP elements ($<10$~eV) have
their abundance ratios relative to Fe consistent with the solar photospheric
ratios, whereas high-FIP elements have their abundance ratios increase
with increasing FIP. We find that the coronal Fe abundance is
consistent with the stellar photospheric value, indicating that
there is no metal depletion in \scrb. However, we obtain a higher Fe 
absolute abundance than in \citet{osten03}. Except for Ar and S, 
our absolute abundances are about 1.5 times larger than those reported by 
\citet{osten03}. However, a comparison of their model with our \xmm\  data (and vice versa) 
shows that both models work adequately in general. We find, therefore, no 
preference for one methodology over the other to derive coronal abundances. 
Despite the systematic discrepancy in absolute abundances, our abundance
ratios are very close to those obtained by \citet{osten03}.
Finally, we confirm the measurement of a low density in \ion{O}{7} 
($< 4 \times 10^{10}$~cm$^{-3}$), but could not confirm the higher densities
measured in spectral lines formed at higher temperatures derived by
other studies of \scrb\  due to the lower spectral resolution of
the \xmm\   grating spectrometers.
\end{abstract}


\section{Introduction}
\label{sect:intro}

Studies of stellar coronae with X-ray missions such as \xmm\  and \chan\ 
have shown peculiar abundance patterns when compared to the 
the coronal composition of the Sun. In the latter, elements with a first
ionization potential (FIP) below about 10~eV appear, on average, overabundant compared to the
high-FIP elements by a factor of a few \citep[e.g.,][]{meyer85}. In addition,
high-FIP elements are of solar photospheric composition
\citep*[e.g.,][]{feldman92,laming95}. This abundance pattern is known as the 
``solar FIP effect''. Previous studies of stellar coronae with lower-resolution
detectors than available onboard with \xmm\  and \chan\  and in the extreme ultraviolet range have
shown a general depletion of the metal abundance \citep[e.g.,][]{antunes94,white94,schmitt96}, and evidence
for a solar-like FIP effect in some stars \citep*{drake97,laming99} or its absence in
others \citep*{drake95}.

\xmm\  and \chan\  have reached unprecedented spectral resolution and
sensitivity in the X-ray regime thanks to their grating spectrometers. 
This led to the discovery of an ``inverse'' FIP effect in HR 1099
\citep{brinkman01,drake01}, in which low-FIP elements are underabundant compared
to the high-FIP elements. Other studies have strengthened the evidence for the presence of
an inverse FIP effect in magnetically active stars
\citep*[e.g.,][]{audard01a,guedel01a,guedel01b,huenemoerder01,huenemoerder03,audard03,
vandenbesselaar03,raassen03,osten03,sanz03a}, 
but have also shown that intermediately active stars show no strong correlation
with the FIP \citep{audard01b}, and less active stars show a solar-like FIP effect \citep{guedel02,telleschi05}.
A transition from the inverse to the solar-like FIP effect thus seems to occur
with decreasing magnetic activity \citep{guedel02,audard03}. However, the inactive
F star Procyon does not show any correlation between its coronal abundances
and the elemental FIP \citep{drake97,raassen02}, although its cool corona ($1-2$~MK)
should display a strong solar-like FIP effect in the transitional picture.
\citet*{sanz04} argued that the comparison of coronal abundances to solar photospheric abundances 
instead of stellar photospheric abundances (which are often unknown or uncertain
in the literature) could be misleading as it could falsely mimic 
inverse or solar-like FIP effects in stars. On the other hand, the photospheric
composition of the solar analogs in the study of \citet{telleschi05} is known to
be similar to the Sun's; therefore, it appears that coronal abundance anomalies
occur, at least in their sample.

Several theoretical studies have produced models to explain the solar FIP effect 
(see reviews by \citealt{henoux95,henoux98}, also \citealt{arge98,schwadron99,mckenzie00}). 
In the stellar context, \citet{guedel02} suggested that 
the non-thermal electrons observed in the radio could explain the inverse FIP effect in active
stars, and 
possibly the FIP effect in inactive stars. Recently, \citet{laming04} proposed a 
model that unifies the FIP and inverse FIP effects by exploring the effects on 
the upper chromospheric plasma of the wave ponderomotive forces. \citet{laming04} 
suggested that the observed solar FIP effect could be turned into an inverse FIP
effect after fine tuning model parameters. Further theoretical works are needed 
to address the wide range of FIP-dependent coronal abundances in the Sun and
stars.
In the mean time, we need to increase the sample of stars
and to study their coronal abundances to understand the mechanism at
the origin of the element fractionation in the Sun, and in stars. In this paper,
we present the \xmm\  study of the RS CVn binary $\sigma^2$ Coronae Borealis which is
composed of two solar analogs as well, albeit spun up by tidal forces.

\section{The RS CV\lowercase{n} Binary $\sigma^2$ Coronae Borealis}
\label{sect:sigcrb}

\objectname{$\sigma^2$ CrB} (HD~146361; TZ CrB) is an active RS CVn binary system 
at a distance of 21.7 pc \citep{perryman97} whose components have spectral types of F9~V and G0~V
\citep{strassmeier03}. \scrb\  is a component of a visual binary with
an orbital period of 1000~years, with the other component ($\sigma^1$ CrB)
at a separation of $6.6\arcsec$. No eclipse takes place in \scrb, due to the low
inclination angle of 28$^{\circ}$ \citep{bakos84}, and the orbit has a
period of $P = 1.14$ days \citep[e.g.,][]{drake89}. \citet{strassmeier03} reported
a $0.017$~d difference between the rotation and orbital periods
and interpreted it as differential rotation of the surface.
The masses and radii of the components are 
close to solar, i.e., $M \sim 1.1 M_{\odot}$ and  $R \sim 1.2 R_{\odot}$
\citep{barden85,drake89}. A detailed list of the physical parameters
of \scrb\   can be found in \citet{strassmeier93} and \citet{strassmeier03}.
 
The binary \scrb\  has been observed extensively from radio to X-rays \citep[see][and
references therein]{osten00}. In the X-ray and EUV regimes, several studies
have investigated the properties of the coronae of \scrb\  
\citep*{agrawal81,agrawal85,agrawal86,vandenoord88,pasquini89,stern92,osten00}. 
Recently, \citet{osten03} presented the \chan\  spectrum of \scrb\  obtained
with the High-Energy Transmission Grating Spectrometer (HETGS) as part of a
coordinated observing campaign in the radio, EUV, and X-ray regimes. They found
a bimodal emission measure distribution with peaks at $6-8$~MK and $30$~MK.
The Fe abundance was found to be depleted as frequently seen in bright
RS CVn binaries \citep{audard03}, whereas Ne/Fe and Ar/Fe abundance ratios
were above solar ratios. Although these ratios are indicative of an inverse FIP effect in
\scrb, other elements did not show a specific correlation with the FIP.
In particular, the low-FIP (6~eV) element Al was overabundant with respect to Fe.
With our analysis of the \xmm\   spectra, we aim to compare derived abundances
with those of \citet{osten03}. We will discuss the coronal abundances of \scrb\  
in the context of the FIP and inverse FIP effects in the Sun and stars.

\section{Observations and Data Reduction}
\label{sect:obsdatared}

\begin{figure}
\includegraphics[width=7.0in]{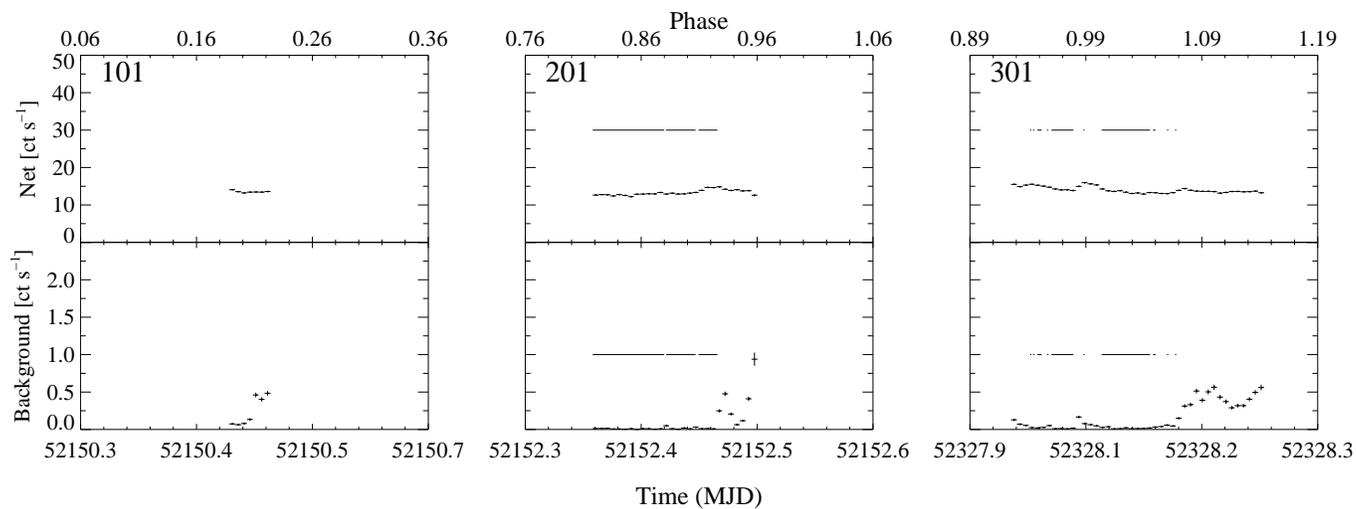}
\caption[lightcurve.eps]{EPIC MOS2 lightcurves with a bin size of 500~s for the three observation dates (error bars
are shown). The selected good time intervals are shown by horizontal lines above the lightcurves. The
excluded time ranges are during high background solar flare activities.
The net count rates are reasonably constant.\label{fig:lc}}
\end{figure}

The observations of \objectname{$\sigma^2$ CrB} were performed by \xmm\   
\citep{jansen01} and were spread over three epochs: 2001 August 29 (hereafter ``101''), 
2001 August 31 (``201''), and 2002 February 22/23 (``301''). The first observation 
lasted for approximately 7~ks and was interrupted by solar flare activity,
therefore we will not discuss this data set further (except as a light curve
in Fig.~\ref{fig:lc}). Solar activity was again high for the two remaining
epochs; nevertheless, we obtained spectra with sufficient quality. We
provide a log of these observations in Table~\ref{tab:log}.

All data were reduced with the \xmm\  Science Analysis System (SAS) 5.4.1 and
the calibration files of June 2003 using standard procedures \citep[e.g.,][]{audard03}. 
The European Photon Imaging Camera (EPIC; \citealt{turner01}) MOS2 data (in SMALL WINDOW mode) were slightly piled-up, and therefore we
excised the center of the point-spread function (PSF) to avoid this problem \citep[e.g.,][]{lumb00}.
A background template was obtained from a region on an outer EPIC MOS2 CCD. 
Its exposure was corrected for vignetting since we assumed here that the dominant
X-ray background was due to the cosmic X-ray background (i.e., the instrumental and particle backgrounds,
which are not vignetted, were negligible).
The EPIC MOS1 data (in TIMING mode) were not used due to the small window size
which prevented us from defining an appropriate background in this mode.
The EPIC pn data were taken in TIMING mode as well and will also not be discussed here.

Despite the high solar activity at the end of each observation (Fig.~\ref{fig:lc}),
we used the complete Reflection Grating Spectrometer (RGS; \citealt{denherder01}) 
data, since a comparison with the ``cleaned'' data (selecting
periods with low solar activity) showed that they were similar.
However, we preferred to use the cleaned EPIC spectra since they showed no significant
contamination at high energies compared to the full-exposure spectra. 
We show the selected good time intervals (GTIs) in Fig.~\ref{fig:lc}.
Thanks to the low count rate variability of \scrb\  during the \xmm\  observations,
the different treatments of GTIs between the EPIC and the RGS data had negligible impact.

In our EPIC images, there was no evidence for X-rays from the visual binary
component $\sigma^1$ CrB. Its X-ray emission could in principle be separated
($6.6\arcsec$) if it was bright enough (indeed, both Castor AB low-mass components
were detected with \xmm\  EPIC MOS for a separation of $3.9\arcsec$; \citealt{guedel01b}). It suggests that
the X-ray luminosity of $\sigma^1$ CrB is much fainter than that of \scrb.
The \chan\   zeroth order image confirms the faintness
the component \citetext{R.~Osten 2003, priv. comm.}. Therefore, no significant
contamination by $\sigma^1$ CrB is expected in the EPIC and RGS data of \scrb.

\section{Data Analysis}
\label{sect:analysis}

The reduced RGS and EPIC MOS2 spectra were fitted simultaneously using collisional
ionization equilibrium (CIE) plasma models with variable elemental abundances
and temperatures. A free constant multiplicative model was allowed to deal with cross-calibration
normalization effects between the RGS and the EPIC spectra. 
We used XSPEC \citep{arnaud96} v11.2.0 software with 
APEC \citep{smith01} code v1.3.0 that contains line and continuum emissivities.
We used the \citet{grev98} set of solar photospheric abundances. Such a set
was also used by \citet{osten03} thus allowing a direct comparison between 
their and our results.

We used a multi-$T$ approach with
free temperatures and emission measures (EMs), and variable abundances 
(C, N, O, Ne, Mg, Si, S, Ar, Ca, Fe). We added a photoelectric absorption component 
using Wisconsin cross-sections of \citet{morrison83} but left the
value of the column density fixed at $2\times10^{18}$~\cmsq\  \citep{osten00} since the
\xmm\  data are not sensitive to such a low value.
\citet{sanz03b} suggest a value of $2.5\times10^{18}$~\cmsq, but the difference
is negligible. We fixed the abundance of Ni to be equal to
the abundance of Fe. 

Furthermore, we used a similar approach as \citet{audard03}. Several wavelength ranges in the 
spectra were discarded since they include low-Z L-shell transitions which are insufficiently
described in current atomic databases \citep{audard01b,audard03,lepson03}. With this procedure, we have
derived the element abundances from the more reliable emission lines of low-Z K-shell ions.
The full list of excluded wavelength ranges is given in Table~\ref{tab:wavrange}.

Figures~\ref{fig:xmmfit1}-\ref{fig:xmmfit2} show the EPIC MOS2 and RGS spectra overlaid with a best fit model
 using a 4-temperature CIE model. The EPIC MOS2 data at wavelengths shorter
than 18~\AA\  were used to model the high-energy tail of the X-ray spectrum and the emission lines inaccessible
to the RGS; significant overlap with the RGS ($8-18$~\AA) was kept to make sure the high-$T$ component
is well linked to the lower-$T$ components which mainly produce emission lines in the RGS band.
The hot Fe K$\alpha$ complex at 6.7~keV (weak in the EPIC MOS2 data but clearly
detected in the EPIC pn data not shown here) indicates the presence of hot plasma
whereas the \ion{O}{7} triplet reveals cool plasma. 
We list the best-fit
parameters of 4-$T$ CIE models for the 201 and 301 observations in Table~\ref{tab:xmmfit}. We also list
the abundances reported by \citet{osten03} for the \chan\  observation of \scrb.

\section{Discussion}

\subsection{Coronal Abundances}
\label{sect:abundances}

We derived absolute abundances (i.e., abundances relative to H) with respect to the
solar photospheric abundances given by \citet{grev98}. 
However, abundance ratios, e.g., relative to Fe, have been found to be more robust \citep[e.g.,][]{audard04,telleschi05}.
Therefore we give abundance ratios to Fe in Table~\ref{tab:xmmfit}. Figure~\ref{fig:abfip} shows the abundance ratios 
as a function of the FIP using the solar photospheric set of \citet{grev98}. We plot the abundance ratios obtained
for the 301 observation shifted by $+0.2$~eV for clarity. The  figure shows two regimes: below 10~eV, low-FIP elements
have about the same abundances and are close to the solar photospheric values. Above 10~eV, the abundance ratios increase with
increasing FIP, as observed in several other RS CVn binaries \citep{audard03}. Figure~11 of \citet{osten03} shows a
similar pattern, although these authors preferred to interpret it as an absence of FIP-dependent pattern.

Figure~\ref{fig:abcmp} shows a comparison of coronal abundances obtained in the best fits to the \xmm\ 
201 and 301 observations, and extends it to a comparison of the abundances obtained by \citet{osten03}.
It shows that abundance ratios are very similar for the three sets. On the other
hand, the absolute abundances are systematically higher in our fits than those by
\citet{osten03} by a factor of about 1.5. Although
two different observations (e.g., at two different epochs) could, in principle, yield different abundances because
the dominant X-ray emitting regions are not the same and the observed, average, coronal composition
could be different, it is far more probable that different analysis techniques could lead to different
best-fit abundances \citep[e.g.,][]{audard04}. Furthermore, the spectral inversion problem being an ill-posed problem, small
statistical variations can lead to discrepant results via the DEM inversion \citep{craig76a,craig76b}. It is, therefore, difficult
to estimate which set of abundances is preferred. On the other hand, it is worthwhile to stress
the robustness of abundance ratios which show (whatever the technique and the data set) that a complex FIP-dependent
abundance pattern is present in \scrb's corona.

The use of a solar photospheric set of abundances to compare with stellar coronal abundances is naturally problematic,
since stars can have a photospheric composition at variance from the Sun's. Magnetically active
stars are difficult
objects to derive accurate surface abundances in view of their rapid rotation and surface spots. 
\citet{strassmeier03} have, however, obtained estimates of the photospheric abundances of the low-FIP elements Fe and Ca,
$A_\mathrm{Fe} =  7.30$ and $A_\mathrm{Ca} = 6.07$ (with the usual $A_\mathrm{H} = 12$ notation; they estimate
an uncertainty of $\sim 0.1$~dex). Comparing
with the solar values from \citet{grev98} ($A_\mathrm{Fe} =  7.50$ and $A_\mathrm{Ca} = 6.36$), \scrb's photosphere 
appears depleted with $\mathrm{[Fe/H] = -0.20}$ and $\mathrm{[Ca/H] = -0.29}$. Our spectral fits obtain
$\mathrm{[Fe/H] \sim -0.15}$ which is consistent with the stellar photospheric value. In contrast, \citet{osten03} obtain
a slightly depleted coronal Fe abundance. 
%In the case of Ca, our abundance estimates are not robust enough, and the \chan\  study did not obtain any Ca abundance. 
It seems safe to conclude that the coronal Fe abundance in \scrb\ 
is close to the photospheric value, in contrast to the solar case where the coronal Fe
abundance is enhanced by a factor
of a few. This situation is also different from that in several magnetically active stars that show a similar U-shape
abundance pattern but a net Fe depletion in their corona \citep[e.g.,][]{guedel01a,sanz03a,telleschi05}. It is difficult to discuss
the case of Ca since we derived large uncertainties and \citet{osten03} did not obtain a Ca abundance.
Nevertheless, a scenario appears to emerge in which low-FIP elements are depleted in very active stars that show
an inverse FIP effect. The abundances of the lowest-FIP elements in intermediately active stars rise up to be quasi-photospheric
while the intermediate-FIP elements still have their abundances depleted, thus the abundance pattern shows a U shape. Finally, in the
less active stars, low-FIP elements have their abundances further increased to become superphotospheric and intermediate-FIP
elements have their abundances quasi-photospheric. Such stars show an abundance pattern similar to the solar FIP effect.

\subsection{A Comparison with the Osten et al. Model}
\label{sect:comparisonosten}
Our methodological approach to the \xmm\  data (i.e., multi-$T$ components) differs significantly from the approach
by \citet{osten03} (i.e., separate treatment of continuum and lines, reconstruction of a differential emission measure
[DEM] distribution based on extracted Fe line fluxes). We
focus this section on comparing their model with ours. Specifically, we test whether their \chan\  model can
provide an adequate fit to our \xmm\  data, and vice versa. Some caveats should be mentioned: i)
\citet{osten03} use the APEC 1.10 version whereas we use the more recent APEC 1.3
version\footnote{http://cxc.harvard.edu/atomdb/index.html for a detailed comparison.}; ii) \scrb's X-ray luminosity
was $1.39$ times higher during the \xmm\  observation ($L_\mathrm{X} = 3.8 \times 10^{30}$~\ergps) than during the \chan\  observation
($L_\mathrm{X} = 2.7 \times 10^{30}$~\ergps) . In our comparison, we
will take this correction factor into account. However, it remains unclear whether this factor should be applied to
all temperature components or only part of them. iii) It is possible that real changes
in the X-ray spectrum (e.g., due to EM or abundance variations) occurred between the two observations. We will
assume in this study that this was not the case.

We have used \citeauthor{osten03}'s quiescent DEM (realization \#2) together with the quiescent abundances 
reported in their Table 5 (we set the C and Ca abundances equal to the Fe abundance, and we set Ni=0.1 times solar
photospheric, as described by \citealt{osten03}). However, after iterations with Dr.~Osten, we discovered that their 
quiescent DEM (as shown in their Fig.~10a) is accurate from $\log T (K) = 6.0$ to $7.4$ for the continuum
and that the last point at $\log T (K) = 7.5$ was used only for the emission lines. However, this temperature
bin essentially contributed a few tens of percent to the flux in the Fe K$\alpha$ complex. This oversight plays 
no significant role for our comparison exercise with the \xmm\  data.
Thus, in practice, we constructed in XSPEC a multi-$T$ model
with 15 components at temperatures at $\log T (K) = 6.0$ to $7.4$ with intervals of $\Delta\log T = 0.1$~dex.
We used \citeauthor{osten03}'s DEM values, multiplied them by the original DEM bin width ($\Delta\log T = 0.1$~dex)
to obtain the total EM for each temperature component, and further divided by 2 since \citet{osten03} accounted for the 
fact that half the radiation is absorbed by the star.%

We have then convolved the model of \citet{osten03} in XSPEC\footnote{\citet{osten03} used the composite trapezoidal rule to 
integrate the line and continuum flux at a specific 
wavelength. The composite trapezoidal rule helps to calculate the integral of a function $f(x)$ over an interval [$a$, $b$] 
subdivided into $N$ subintervals [$x_i$, $x_{i+1}$] of widths $h = (b - a) / N$ (thus $x_i = a + i \times h$ for 
$i = 0, 1, \dots, N$).  The integral is then equal to $h \times (f(a) + f(b)) / 2 + h \times \sum_{i=1}^{N-1} f(x_i)$.
Thus, we used half the DEM values at $\log T (K) = 6.0$ and $7.4$ in XSPEC.}  (using our RGS RMF file); we have also convolved our
\xmm\  EPIC MOS2 (301) model through the ACIS-S/HETGS first-order RMF/ARF provided by Dr.~Osten\footnote{We have
checked that our 15-$T$ model based on \citeauthor{osten03}'s model was correct by convolving it through the same
\chan\  response matrix files and comparing it with their synthesized model provided by Dr.~Osten.}.
Figures~\ref{fig:xmmwchan1}-\ref{fig:xmmwchan3} compare the \xmm\  RGS data overlaid with \citeauthor{osten03}'s \chan\  model, whereas 
Figures~\ref{fig:chanwxmm1}-\ref{fig:chanwxmm3} show the \chan\  HETGS data (provided by Dr.~Osten) overlaid with the \xmm\  model.

We find surprisingly good agreement, despite different absolute abundances. There are, however, some
discrepancies. In Fig~\ref{fig:xmmwchan2}, the \ion{Fe}{17} $\lambda 15$~\AA\  line is overestimated compared to the \xmm\  data. But this is
no different from the analysis of \citet{osten03} which overestimated the \chan\  \ion{Fe}{17} line flux as well (see their
Fig.~10a, center panel). In addition, the \citeauthor{osten03} model overestimates the RGS continuum level
by about 40\% at wavelengths longer than 25~\AA\  (Fig.~\ref{fig:xmmwchan3}). A few
possible explanations can be put forward: i) the \chan\  HETGS wavelength range ends around 25~\AA, therefore the 
\citeauthor{osten03}  model had no possibility to accurately take into account the flux beyond this instrumental limit.
\citet{sako03} reported a similar situation in which the model of \citet{lee01}, based on HETGS data, could reproduce
the general properties of the RGS spectrum of MCG -6-30-15 below 23~\AA, but it overestimated the flux for $\lambda >
25$~\AA; ii) calibration
problems at long wavelengths are possible; however, the 40\% overestimate is larger than the current
calibration knowledge; iii) it was inadequate to multiply 
the EMs of \citeauthor{osten03}'s model by a constant value of $1.39$ to account for the X-ray luminosity difference of \scrb\  between the \chan\  and 
\xmm\  observations. It is possible that the variation is due to some part of the DEM only (e.g., at
high temperatures). However, the long-wavelength continuum
level is essentially determined by the high-temperature components (which describe the short and mid-wavelength ranges accurately), 
and, therefore we find this possibility less probable. Instead we propose that the lack of spectral coverage in the range
$25-40$~\AA\  by \textit{Chandra} HETGS slightly biased the EMD of
\citet{osten03} to overestimate the long-wavelength
flux. We note that, although \citet{osten03} used \textit{EUVE} line fluxes to determine their DEM, these authors did not 
make use of the continuum level in the \textit{EUVE} wavelength range (which is about 5 times lower than the continuum 
level at 30~\AA). 

Our \xmm\  model (whose EMs were divided by $1.39$) shows generally excellent agreement
with the \chan\ data (Fig.~\ref{fig:chanwxmm1}-\ref{fig:chanwxmm3}). However, the flux in the 
\ion{Ar}{17} triplet near $4$~\AA\ was underestimated; it suggests that Ar abundances obtained
from simultaneous EPIC+RGS fits might be underestimated generally. The \citeauthor{osten03} model, however, 
predicts L-shell fluxes of Ar and S lines at long wavelengths which are consistent with the RGS data 
(Fig.~\ref{fig:xmmwchan3}), despite the long-standing problem of the lack of atomic data of L-shell lines of low-Z elements 
or their inaccuracy in current atomic databases  \citep{audard01b,raassen02,lepson03}. 
%It is unclear whether inaccurate emissivities in APEC 1.10 
%(used by \citealt{osten03} and in this comparison) is accountable for this discrepancy. 
%The latter is unlikely since the \citeauthor{osten03} model
%was obtained by a simultaneous treatment of quiescent \textit{EUVE} and \chan\  spectra. Indeed the lack of data of L-shell lines
%of low-Z elements or their inaccuracy in current atomic databases is a long-standing problem \citep{audard01b,raassen02,lepson03}.

With a few exceptions, we find, therefore, excellent agreement among different models derived with different techniques from different
data sets, but which show very similar abundance ratios but different absolute abundances. This exercise casts some doubt on the
preference of one method over the other to determine coronal abundances, and it demonstrates that the determination of true
absolute abundances remains a challenge.\footnote{Although not shown in this paper, we note that setting the absolute Fe abundance to
solar photospheric and using the same abundance ratios as provided by \citet{osten03}, it is impossible to obtain a reasonably good 
fit of the \chan\  data by simply multiplying the EMs by a constant factor. If the model is adjusted to match the continuum level, 
the Fe line fluxes are indeed systematically overestimated. It again reinforces the result that Fe is underabundant in \scrb's corona compared to
the solar photospheric Fe abundance.} 
Other studies support this interpretation. For example, \citet{audard04} studied the bright FK Com-type giant YY Men using
different methods and found that, for a given data set (either \chan\  or \xmm), slight differences in absolute abundances were derived. 
However, they found excellent agreement for the abundance ratios relative to Fe. In parallel, \citet{telleschi05} studied the 
\xmm\  data of a sample of solar analogs with different methods as well. They also observed the robustness of abundance ratios 
and demonstrated the presence of a transition from a solar-like FIP
effect in inactive stars to an inverse FIP effect in active stars. \citet{garcia05} used one method but different sets of emission 
lines to derive the coronal abundances in AB Dor (and V471 Tau) and found similar abundances. They further compared their results with those 
in earlier studies and found good agreement in the general trend, albeit with some significant differences. Thus, in brief, despite the
use of various methodologies, abundance patterns as a function of the FIP appear robust and look dependent on the magnetic activity level.
However, it remains challenging i) to constrain absolute abundances, and ii) to compare coronal abundances to their photospheric
counterparts.

\subsection{Electron Density}
\label{sect:densities}

High-resolution X-ray spectra observed with \xmm\  and \chan\  allow us to determine electron densities
from line ratios of He-like ions. The ratio $R$ of the flux in the forbidden ($f$) line to the flux in the intercombination ($i$)
line is indeed density-sensitive. In particular, the \ion{O}{7} triplet is most useful in measuring
coronal densities of the order of $10^{10}$~\cmcc, and is relatively blend-free (in contrast to \ion{Ne}{9}),
and well-resolved (in contrast to, e.g., \ion{Mg}{11} and \ion{Si}{13} for the RGS). We have obtained line fluxes
for the \ion{O}{7} triplet using delta functions convolved through the instrumental response.
We list them in Table~\ref{table4} together with 68\% confidence ranges ($\Delta\chi^2 = 1$).
The $R$ ratio is equal to $2.81 \pm 0.77$, corresponding to an upper limit to the electron density of $4 \times 10^{10}$~\cmcc\ 
\citep{porquet01}, assuming a photoexcitation contribution by the radiation field with $T_\mathrm{eff} \sim 6,000$~K.

\citet{osten03} and \citet{testa04} reported for \ion{O}{7} %$R= 2.33 \pm 0.78$, i.e., consistent with our $R$ value. Furthermore,
an electron density of $\sim 2 \times 10^{10}$~\cmcc, i.e., consistent with ours. These authors and \citet{sanz03b} quote 
(higher) electron densities from other He-like triplets and Fe lines (in the EUV range) which we cannot confirm due to the poorer 
spectral resolution of the RGS (or the lack of wavelength coverage). %%We obtain line ratios consistent with the low-density limits.
%However, a comparison of the \ion{Mg}{11} and 
%\ion{Ne}{9} triplets with the model of \citet{osten03} (calculated for an electron density of $n_\mathrm{e} = 1$~\cmcc) 
%does not show an indication for significantly higher densities than those bounded by the low-density limits (Fig.~\ref{fig:xmmwchan1}).
Unfortunately, the \ion{N}{6} triplet is too faint to measure any line flux.

\subsection{Opacity}
\label{sect:opacity}

Our CIE models  assume an optically thin plasma, so it is
necessary to verify that opacity effects play a negligible role. We examined the
ratio of the Fe \textsc{xvii} spectral lines at 15.01~\AA\   and at 15.26~\AA, since
this ratio is sensitive to opacity effects (the 15.01~\AA\  line has a large oscillator strength).
The measured Fe \textsc{xvii} fluxes are shown in Table \ref{table4}.

Until recently, theoretical values for the ratio used to range from $3.0$ to $4.7$ \citep{bhatia92},  
significantly different from laboratory measurements which suggest a range from $2.8$ to $3.2$ \citep{brown98,laming00,brown01}.
However, recent theoretical work has brought the theoretical ratio closer to the laboratory ratio \citep{chen02,chen03,fournier05}.
We obtain a flux ratio in \scrb\  of $3.02 \pm 0.15$, i.e.,  consistent with the the ratios derived by \citet{osten03}
($3.01 \pm 0.26$ with HEG and $2.94 \pm 0.15$ with MEG). \citet{ness03} derived a slightly lower ratio ($2.56 \pm 0.13$), 
probably because of a different treatment of the continuum. Nevertheless, they
also concluded that this value was close to the ratio obtained in laboratory 
measurements. Thus, the line ratio indicates that there are no significant 
opacity effects in the coronae of \scrb.

\section{Summary and Conclusions}
\label{sect:conclusions}

We have presented a study of the \xmm\  observations of the binary \scrb. No large flare was detected during the
observations.
We focused our analysis on the determination of the coronal abundances
by means of a multi-$T$ approach. Although such models do not give an accurate representation of the coronal DEM,
they suffice to obtain good measurements of coronal abundances and abundance ratios in particular \citep[e.g.,][]{audard04,telleschi05}.
We have, furthermore, compared our resulting abundances with those of \citet{osten03} who published the \chan\  data of \scrb.
We find very good agreement between their abundance ratios (relative to the Fe abundance) and ours, but find a systematic difference
in the absolute abundances. Upon checking whether their model could fit our \xmm\  data, we have obtained an excellent
agreement (after correcting the published DEM, see Section \ref{sect:comparisonosten}). At wavelengths longer than 25~\AA, their model 
overestimates the continuum level. We argue that their model, based on HETGS data below this wavelength,
did not constrain their DEM well enough to account for the longer wavelength range. Our \xmm\  best-fit model
also reproduces the \chan\  HETGS spectra remarkably well. Thus, abundance ratios are found to be robust, whereas the discrepancy in absolute
abundances found between our model and \citeauthor{osten03}'s model probably stems from the different methodologies. Similar results
have been found elsewhere \citep{audard04,telleschi05,garcia05}. Nevertheless the presence of abundance patterns as a function of
the FIP seems reliable.

The abundance ratios relative to the coronal Fe abundance showed a correlation with the FIP, but such a correlation was neither similar to the inverse FIP
effect observed in very active binaries \citep[e.g.][]{audard03} nor to the solar-like FIP effect observed in less active 
stars \citep[e.g.][]{telleschi05}. The abundance ratios of low-FIP elements ($< 10$~eV) with respect to Fe are similar to the solar photospheric
abundance ratios; for intermediate-FIP elements, they are however considerably lower but increase with increasing FIP. 
Such an abundance pattern has already been observed in other magnetically active stars \citep{guedel01a,sanz03a,huenemoerder03}. However, \citet{garcia05}
found the abundances of the very low-FIP elements (Al, Ca, Na) in AB Dor to be similar to, or lower than, those of low FIP (Mg, Si, Fe),
in contrast to \citeauthor{sanz03a}'s results. It is possible that
these patterns are real, but we emphasize that coronal abundances are often compared to the solar photospheric composition 
due to the lack of available or reliable measurements in magnetically active, fast-rotating stars. In the case of \scrb,
the Fe and Ca photospheric abundances of the \scrb\  binary were available \citep{strassmeier03} and suggested that the
low Fe abundance in \scrb's coronae was in fact consistent with the photospheric value. 

Finally, we obtained an upper limit of the coronal density based on the \ion{O}{7} triplet ($n_\mathrm{e} < 4 \times 10^{10}$~\cmcc)
which is consistent with other measurements in \scrb\  \citep{osten03,testa04}. 
\citet{osten03}, \citet{sanz03b}, and \citet{testa04} found higher densities ($n_\mathrm{e} \sim 10^{12}$~\cmcc) from spectral
lines formed at higher temperatures than \ion{O}{7}. However, due to the lower spectral resolution of the RGS  (compared to \chan\  HETGS), 
we could not confirm such coronal densities.

In conclusion, this study has confirmed that the abundance ratio pattern in \scrb's coronae does not follow that of
a simple inverse FIP effect or of a solar-like FIP effect, but appears to be more complex. However, the lack of data on
photospheric abundances in \scrb\  to compare with still casts doubt on whether such a pattern is real or mirrors the stellar
composition. Nevertheless, the underabundance of Fe in \scrb's coronae (compared to the solar photospheric Fe abundance)
appears robust and is consistent with its abundance found in the stellar photosphere.

\acknowledgments

Dr.~Rachel Osten is warmly thanked for providing  the \chan\  data and model and access to
her software, and for useful comments that improved the content of this paper. 
The Columbia group acknowledges support from NASA to Columbia University for 
\textit{XMM-Newton} mission support and data analysis. MG acknowledges support from the Swiss 
National Science Foundation  (grant 20-66875.01).
Based on observations obtained with \xmm, an ESA science 
mission with instruments and contributions directly funded by 
ESA Member States and the USA (NASA). We thank an anonymous referee for useful comments
that improved this paper.

\clearpage

\begin{deluxetable}{lllllcc}
\tablecaption{Observation Log.\label{tab:log}}
\tabletypesize{\scriptsize}
\tablewidth{0pt}
\tablehead{ 
 &  & & \colhead{Observation Start} & \colhead{Observation
End} & \colhead{Total Exposure} & \colhead{Filtered Exposure\tablenotemark{a}}\\
\colhead{Instrument} & \colhead{Mode} & \colhead{Filter} & \colhead{(UT)} & \colhead{(UT)}
& \colhead{(ks)} & \colhead{(ks)}}

\startdata
\multicolumn{7}{c}{\textsc{Revolution 316, Observation ID 0111470201 (``201'')}}\\
\vspace{-2mm}
\\
RGS1\hspace{0.2mm}\dotfill\  & Spectroscopy & None & 2001 Aug 31 07:38:02 & 2001
Aug 31 13:02:20 & 19.5 & 18.9 \\
RGS2\hspace{0.2mm}\dotfill\ & Spectroscopy & None & 2001 Aug 31 07:38:02 & 2001 Aug
31 13:05:11 & 19.6 & 18.5\\
MOS1\tablenotemark{b}\hspace{-0.4mm}\dotfill\ & TIMING & Thick & 2001 Aug 31 07:44:30 & 2001 Aug 31 11:32:36 &
13.7 & \nodata \\
MOS2\hspace{-0.4mm}\dotfill\ & IMAGING & Medium & 2001 Aug 31 07:44:33 & 2001 Aug 31
11:34:12 & 13.8 & 9.8\\
PN\hspace{-0.4mm}\dotfill\ & TIMING & Medium & 2001 Aug 31 08:13:58 & 2001 Aug 31
13:08:00 & 17.6 & \nodata\\
\hline
\vspace{-2mm}
\\

\multicolumn{7}{c}{\textsc{Revolution 404, Observation ID 0111470301 (``301'')}}\\
\vspace{-2mm}
\\
RGS1\hspace{0.2mm}\dotfill\ & Spectroscopy & None & 2002 Feb 22 23:23:29 & 2002 Feb
23 05:47:04 & 23.0 & 22.5\\
RGS2\hspace{0.2mm}\dotfill\ & Spectroscopy & None & 2002 Feb 22 23:23:29 & 2002 Feb
23 05:47:05 & 23.0 & 21.8\\
MOS1\tablenotemark{b}\hspace{-0.4mm}\dotfill\ & TIMING & Thick & 2002 Feb 22 23:29:57 & 2002 Feb 23
05:27:24 & 21.4 & \nodata \\
MOS2\hspace{-0.4mm}\dotfill\ & IMAGING & Medium & 2002 Feb 22 23:29:59 & 2002 Feb 23
05:27:53 & 21.5 & 6.7\\
PN\tablenotemark{b}\hspace{-0.4mm}\dotfill\ & TIMING & Medium & 2002 Feb 22 23:59:25 & 2002 Feb 23
05:46:27 & 20.8 & \nodata\\
\vspace{-2mm}

\enddata
\tablecomments{Data from Revolution 315 are not included for analysis due to short
exposure times of approximately 7ks and significant background contamination from
solar flare activity.}
\tablenotetext{a}{Maximum effective integration time used for data analysis.}
\tablenotetext{b}{The EPIC MOS1 and pn data were not used in this paper.}
\end{deluxetable}

\clearpage

%%%%%%%%%%%
\begin{deluxetable}{ll}
\tabletypesize{\footnotesize}
\tablecaption{Wavelength Ranges Excluded from the Fitting Procedure.\label{tab:wavrange}}
\tablewidth{0pt}
\tablehead{ 
 \colhead{Instrument} & \colhead{$\lambda$ Range}\\
 & \colhead{(\AA)}}
 \startdata
RGS\dotfill & $\leq$ 8.3\\
RGS\dotfill & 15.80-16.20\\
RGS\dotfill & 18.60-18.80\\
RGS\dotfill & 19.50-19.90\\
RGS\dotfill & 20.00-20.30\\
RGS\dotfill & 20.60-20.80\\
RGS\dotfill & 21.15-21.30\\
RGS\dotfill & 21.40-21.50\\
RGS\dotfill & 23.40-23.70\\
RGS\dotfill & 24.10-24.60\\
RGS\dotfill & 24.90-25.25\\
RGS\dotfill & 25.45-25.80\\
RGS\dotfill & 26.25-26.50\\
RGS\dotfill & 27.00-27.20\\
RGS\dotfill & 27.30-27.70\\
RGS\dotfill & 27.80-28.10\\
RGS\dotfill & 28.30-28.50\\
RGS\dotfill & 29.40-29.70\\
RGS\dotfill & 30.20-30.60\\
RGS\dotfill & 30.90-31.15\\
RGS\dotfill & 31.50-31.70\\
RGS\dotfill & 32.10-33.55\\
RGS\dotfill & 33.85-34.00\\
RGS\dotfill & 34.50-34.65\\
RGS\dotfill & 35.60-35.80\\
RGS\dotfill & 36.30-36.85\\
RGS\dotfill & 37.50-37.80\\
MOS2\dotfill & $\leq$ 1.65\\
MOS2\dotfill & $\geq$ 18\\
\enddata

\end{deluxetable}

\clearpage

%%%%%%
\begin{deluxetable}{lrrr}
\tabletypesize{\footnotesize}
\tablecolumns{4}
\tablewidth{0pc}
\tablecaption{Best-Fit Parameters for RGS+MOS2 Together with Abundances from \citet{osten03}.\label{tab:xmmfit}}
\tablehead{
\colhead{\makebox[3cm][l]{Parameters}} & \colhead{201}&  \colhead{301} & \colhead{Osten et al. (2003)}}
\startdata
$\mathrm{[C/Fe]}$\dotfill &  $-0.24_{-0.10}^{+0.05}$ &   $-0.28_{-0.09}^{+0.08}$	& \nodata \\
$\mathrm{[N/Fe]}$\dotfill &  $-0.09_{-0.11}^{+0.08}$ &   $-0.16_{-0.10}^{+0.10}$	& $-0.07_{-0.08}^{+0.07}$\\
$\mathrm{[O/Fe]}$\dotfill &  $-0.22_{-0.04}^{+0.02}$ &   $-0.22_{-0.05}^{+0.03}$   & $-0.16_{-0.01}^{+0.01}$\\
$\mathrm{[Ne/Fe]}$\dotfill & $+0.11_{-0.03}^{+0.04}$ &   $+0.15_{-0.05}^{+0.05}$  & $+0.16_{-0.01}^{+0.01}$\\
$\mathrm{[Mg/Fe]}$\dotfill & $+0.04_{-0.04}^{+0.03}$ &   $+0.08_{-0.05}^{+0.05}$   & $-0.00_{-0.01}^{+0.01}$\\
$\mathrm{[Si/Fe]}$\dotfill & $-0.06_{-0.07}^{+0.04}$ &   $-0.09_{-0.09}^{+0.08}$  & $-0.10_{-0.01}^{+0.01}$\\
$\mathrm{[S/Fe]}$\dotfill &  $-0.45_{-0.19}^{+0.13}$ &   $-0.49_{-0.32}^{+0.19}$  & $-0.28_{-0.05}^{+0.05}$\\
$\mathrm{[Ar/Fe]}$\dotfill & $+0.04_{-0.35}^{+0.29}$ &   $-0.35_{-\infty}^{+0.52}$	    & $+0.34_{-0.14}^{+0.11}$\\
$\mathrm{[Ca/Fe]}$\dotfill & $-0.12_{-\infty}^{+0.30}$ & $+0.16_{-0.53}^{+0.25}$  & \nodata\\
$\mathrm{[Fe/H]}$\dotfill &  $-0.17_{-0.01}^{+0.01}$ &    $-0.13_{-0.02}^{+0.02}$ & $-0.34_{-0.02}^{+0.02}$\\
$T_{1}$ (MK)\dotfill & $3.83_{-0.26}^{+0.10}$ & $3.34_{-0.19}^{+0.18}$  & \nodata\\
$T_{2}$ (MK)\dotfill & $7.74_{-0.08}^{+0.08}$ & $7.48_{-0.08}^{+0.08}$  & \nodata\\
$T_{3}$ (MK)\dotfill & $14.7_{-0.52}^{+0.56}$ & $14.4_{-0.60}^{+0.50}$  & \nodata\\
$T_{4}$ (MK)\dotfill & $28.4_{-0.13}^{+8.19}$ & $31.6_{-3.40}^{+5.00}$  & \nodata\\
$\log \mathrm{EM}_{1}$ (\cmcc)\dotfill & $52.7_{-0.10}^{+0.10}$ & $52.5_{-0.10}^{+0.10}$	& \nodata\\
$\log \mathrm{EM}_{2}$ (\cmcc)\dotfill & $53.1_{-0.10}^{+0.10}$ & $53.1_{-0.10}^{+0.10}$	& \nodata\\
$\log \mathrm{EM}_{3}$ (\cmcc)\dotfill & $52.9_{-0.10}^{+0.10}$ & $52.9_{-0.10}^{+0.10}$	& \nodata\\
$\log \mathrm{EM}_{4}$ (\cmcc)\dotfill & $52.7_{-0.10}^{+0.10}$ & $52.8_{-0.10}^{+0.10}$	& \nodata\\
$\chi^2$\dotfill & 1.34 &1.48\\
\enddata

\end{deluxetable}

\clearpage

%%%%
\begin{deluxetable}{lc}
\tabletypesize{\small}
\tablecaption{Photon Fluxes for O \textsc{vii} Triplet and Fe
\textsc{xvii} Lines.\label{table4}}
\tablewidth{0pt}
\tablehead{ 
\colhead{Wavelength} & \colhead{Flux} \\
\colhead{(\AA)} & \colhead{($10^{-4}$ \flx)} }
\startdata
21.60\dotfill& 3.65 $\pm$ 0.28\\
21.80\dotfill& 0.80 $\pm$ 0.20\\
22.10\dotfill& 2.25 $\pm$ 0.25\\
15.01\dotfill& 13.9 $\pm$ 0.30\\
15.26\dotfill& 4.60 $\pm$ 0.21
\enddata

\end{deluxetable}

\clearpage

%%%%%%%%%%%%%%%%%%%%%%%%%%%%%%%%%%%%%%%%%%%%%%%%%%%%%%%%%%%%%%%%%%%%%%%%%%%%%%

\clearpage
%%%%%%%%%%%%%%%%%%%%%%%%%%%%%%%%%%%%%%%%%%%%%%%%%%%%%%%%%%%%%%%%%%%%%%%%%%%%%%
\renewcommand{\thefigure}{\arabic{figure}\alph{subfigure}}
\setcounter{subfigure}{1}
\begin{figure}
\includegraphics[width=7.0in]{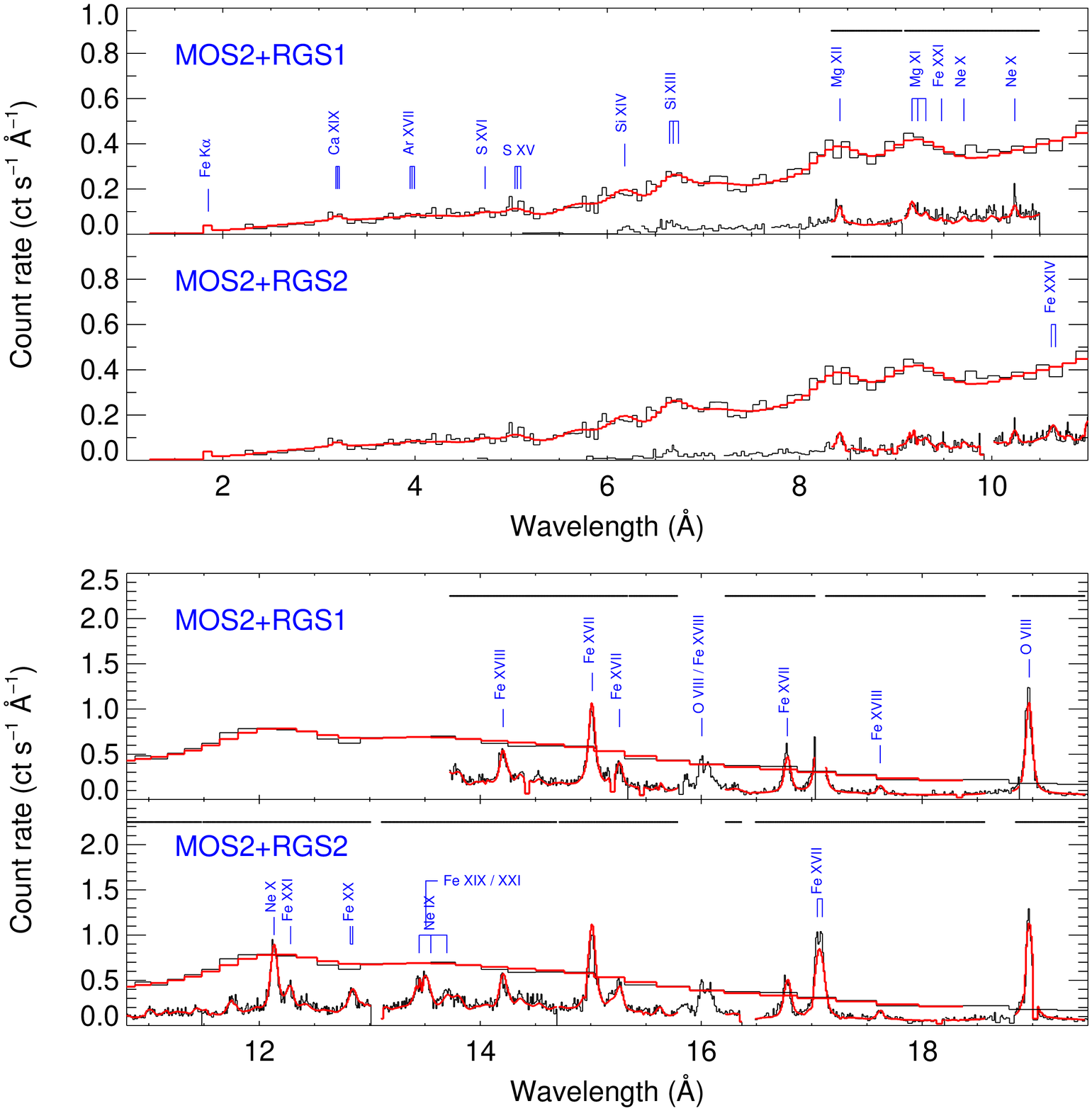}
\caption[xmm1.eps]{($a$) \xmm\  MOS2, RGS1, and RGS2 spectra with the EPIC MOS2+RGS 
(301; see Tab.~\ref{tab:xmmfit}) best-fit model (superposed line) for $1-20$~\AA. 
Error bars are not shown for clarity.  The model is only shown in the wavelength ranges used
for the spectral fits (see Tab.~\ref{tab:wavrange}). The ranges for the RGS are also shown 
at the top of each panel as broken lines. ($b$) Same as
($a$), but for $20-40$~\AA.\label{fig:xmmfit1}}
\end{figure}

\clearpage

%%%%%%%%%%%%%%%%%%%%%%%%%%%%%%%%%%%%%%%%%%%%%%%%%%%%%%%%%%%%%%%%%%%%%%%%%%%%%%
\addtocounter{figure}{-1}
\addtocounter{subfigure}{1}
\begin{figure}
\includegraphics[width=7.0in]{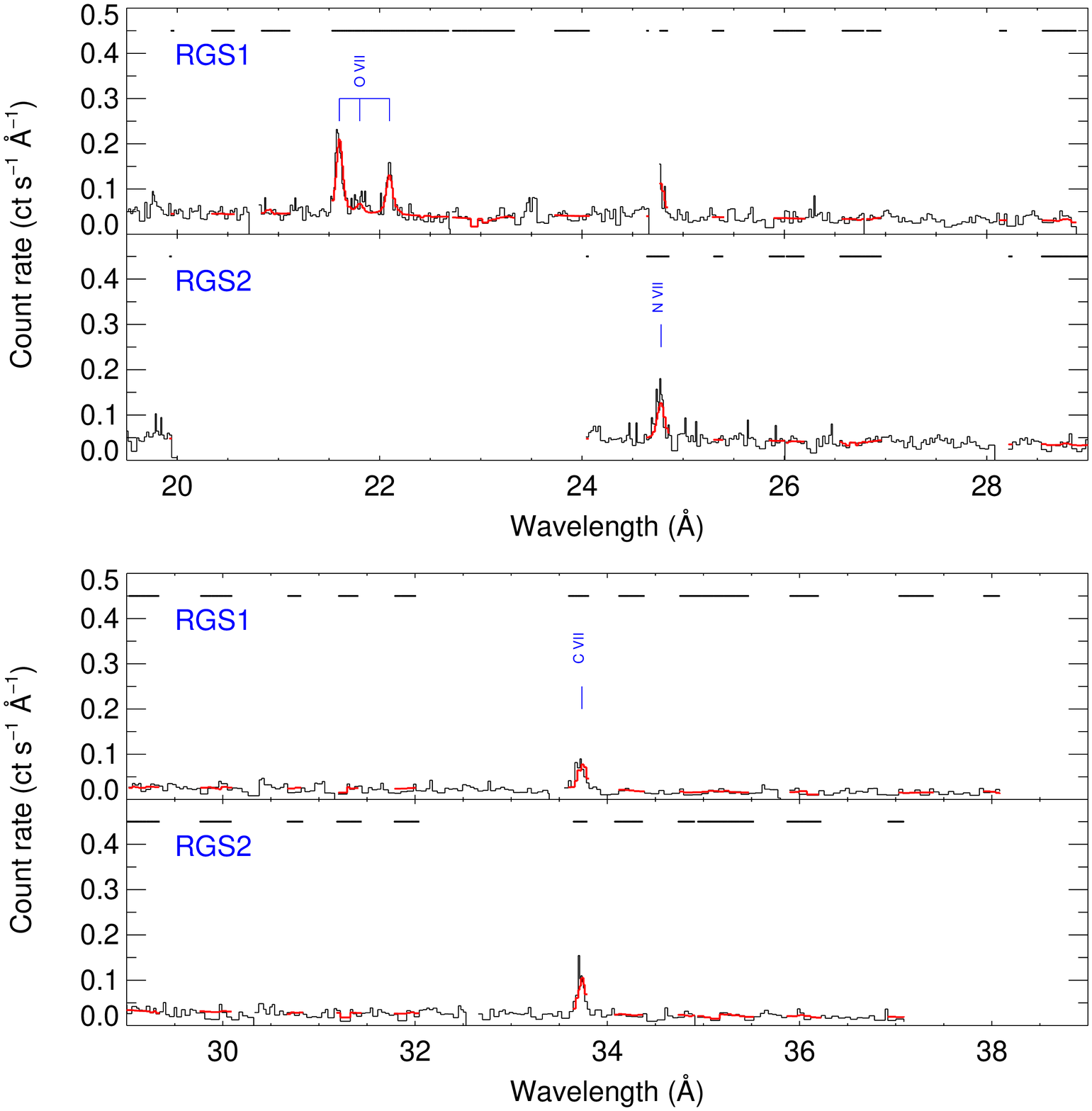}
\caption[xmm2.eps]{\label{fig:xmmfit2}}
\end{figure}

\clearpage

%%%%%%%%%%%%%%%%%%%%%%%%%%%%%%%%%%%%%%%%%%%%%%%%%%%%%%%%%%%%%%%%%%%%%%%%%%%%%%
\renewcommand{\thefigure}{\arabic{figure}}
\begin{figure}
\includegraphics[width=7.0in]{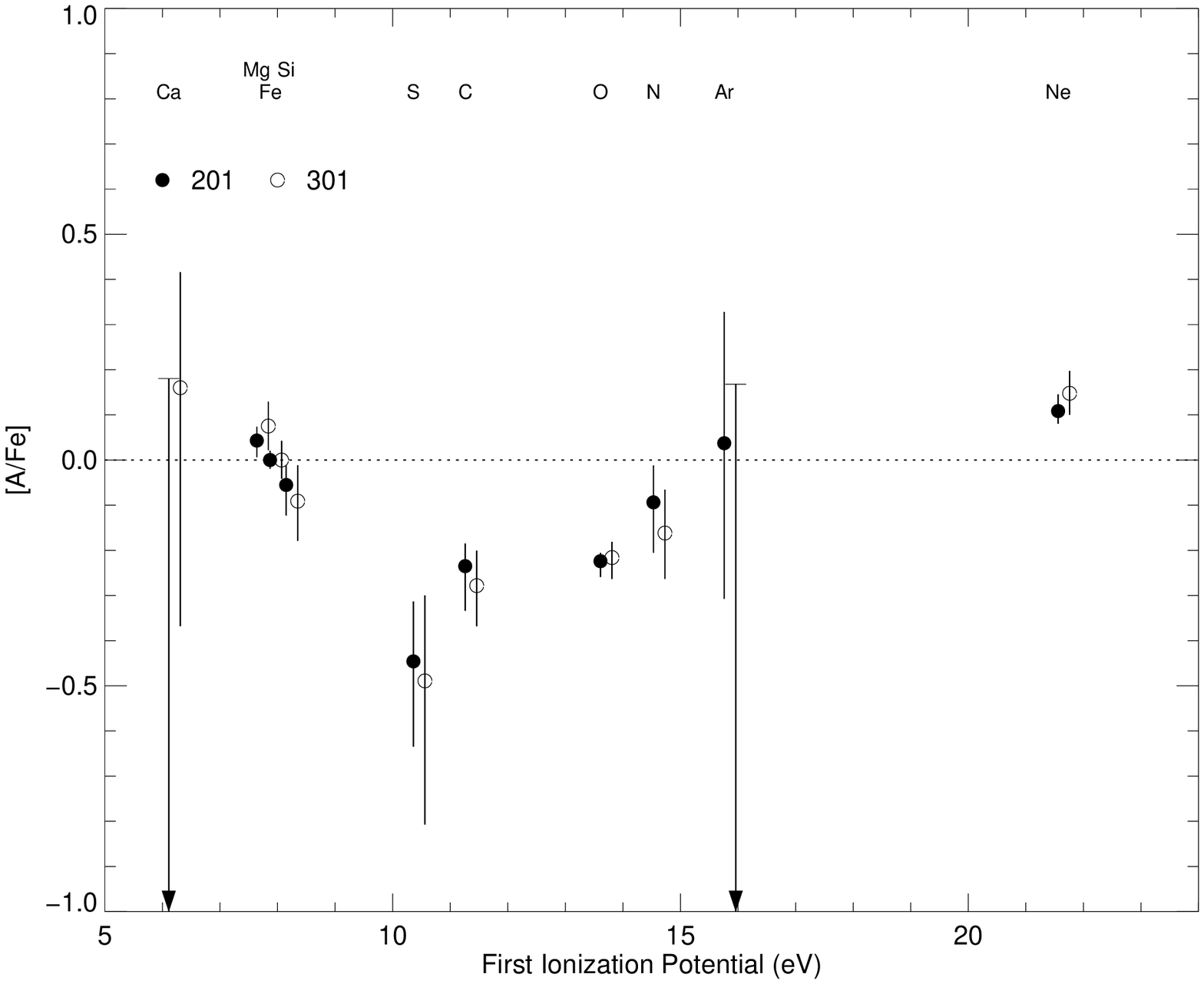}
\caption{Abundance ratios relative to Fe and relative to solar abundances \citep{grev98} for the 201 and 301 observations as a
function of the FIP. For clarity, we shifted the 301 data points by $+0.2$~eV. Statistical errors only are shown; 
systematic errors probably range up to $0.1-0.2$~dex. For Ca (201) and Ar (301), upper limits are shown with
an arrow down to the bottom of the figure.\label{fig:abfip}}
\end{figure}

\clearpage

%%%%%%%%%%%%%%%%%%%%%%%%%%%%%%%%%%%%%%%%%%%%%%%%%%%%%%%%%%%%%%%%%%%%%%%%%%%%%%
\begin{figure}
\includegraphics[width=7.0in]{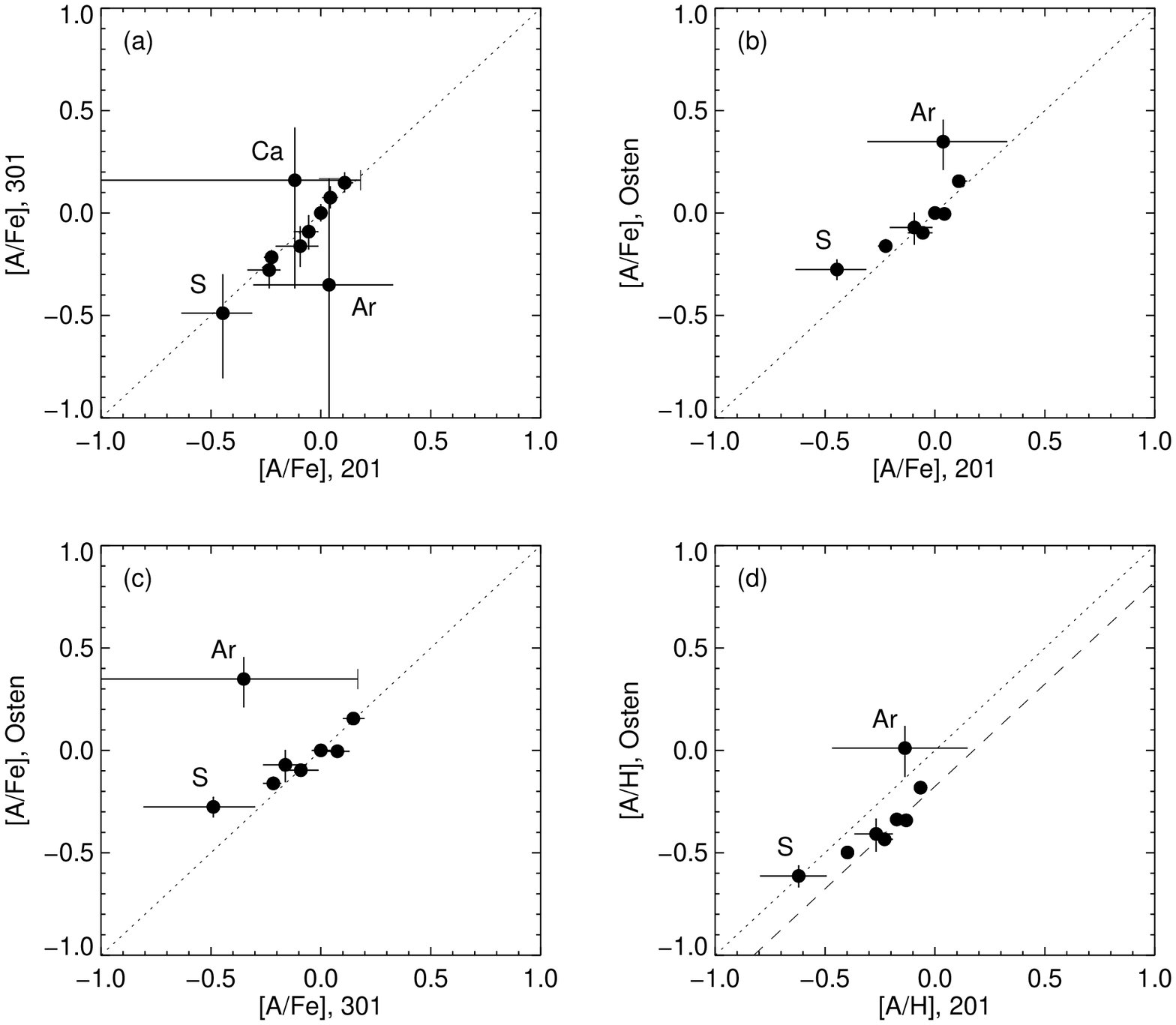}
\caption{Comparisons between abundances obtained in this paper and with those by \citet{osten03}. The bottom right panel (d) shows
absolute abundances instead of abundance ratios to Fe (a,b,c). Abundance ratios are robust and similar, whereas our
absolute (relative to H) abundances are systematically higher than those derived by \citet{osten03}. Some abundances discussed in the
text are labeled for clarity. The dotted line represents a 1:1 correlation,
whereas the dashed line in panel d represents a 1.5:1 correlation.\label{fig:abcmp}}
\end{figure}

\clearpage

%%%%%%%%%%%%%%%%%%%%%%%%%%%%%%%%%%%%%%%%%%%%%%%%%%%%%%%%%%%%%%%%%%%%%%%%%%%%%%
\renewcommand{\thefigure}{\arabic{figure}\alph{subfigure}}
\setcounter{subfigure}{1}
\begin{figure}
\centering
\includegraphics[width=\textwidth]{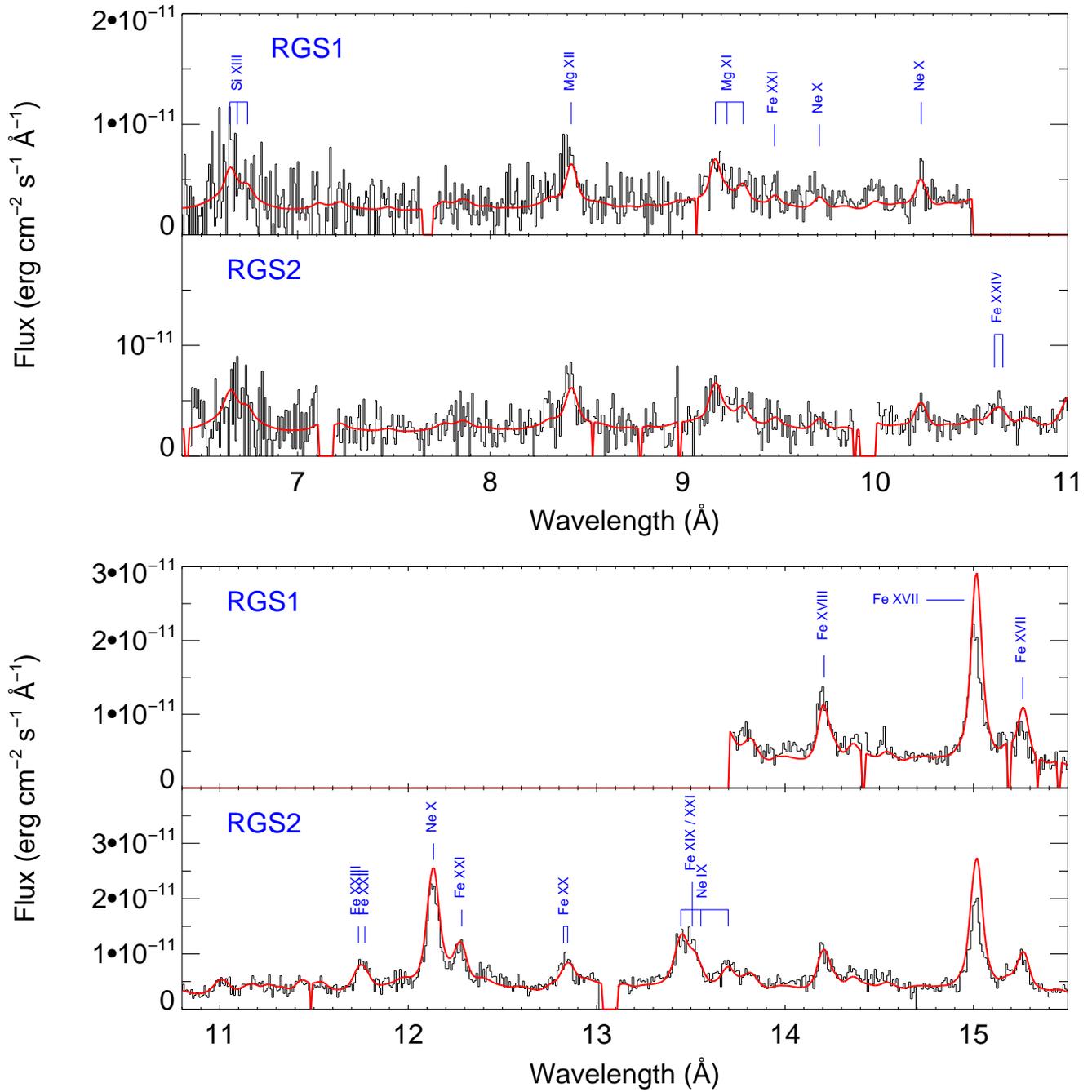}
\caption{($a$) \xmm\  RGS1 and RGS2 fluxed spectra (uncorrected for interstellar absorption which is in any case negligible in
this wavelength range) compared with the \chan\  quiescent model 
(superposed line) of \citet{osten03} for $6.5-15.5$~\AA. Error bars are not shown for clarity. 
($b$) Same as ($a$), but for $15-25$~\AA. ($c$) Same as ($a$) but for $25-35$~\AA.\label{fig:xmmwchan1}}
\end{figure}
%\newpage

\clearpage

%%%%%%%%%%%%%%%%%%%%%%%%%%%%%%%%%%%%%%%%%%%%%%%%%%%%%%%%%%%%%%%%%%%%%%%%%%%%%%
\addtocounter{figure}{-1}
\addtocounter{subfigure}{1}
\begin{figure}
\centering
\includegraphics[width=\textwidth]{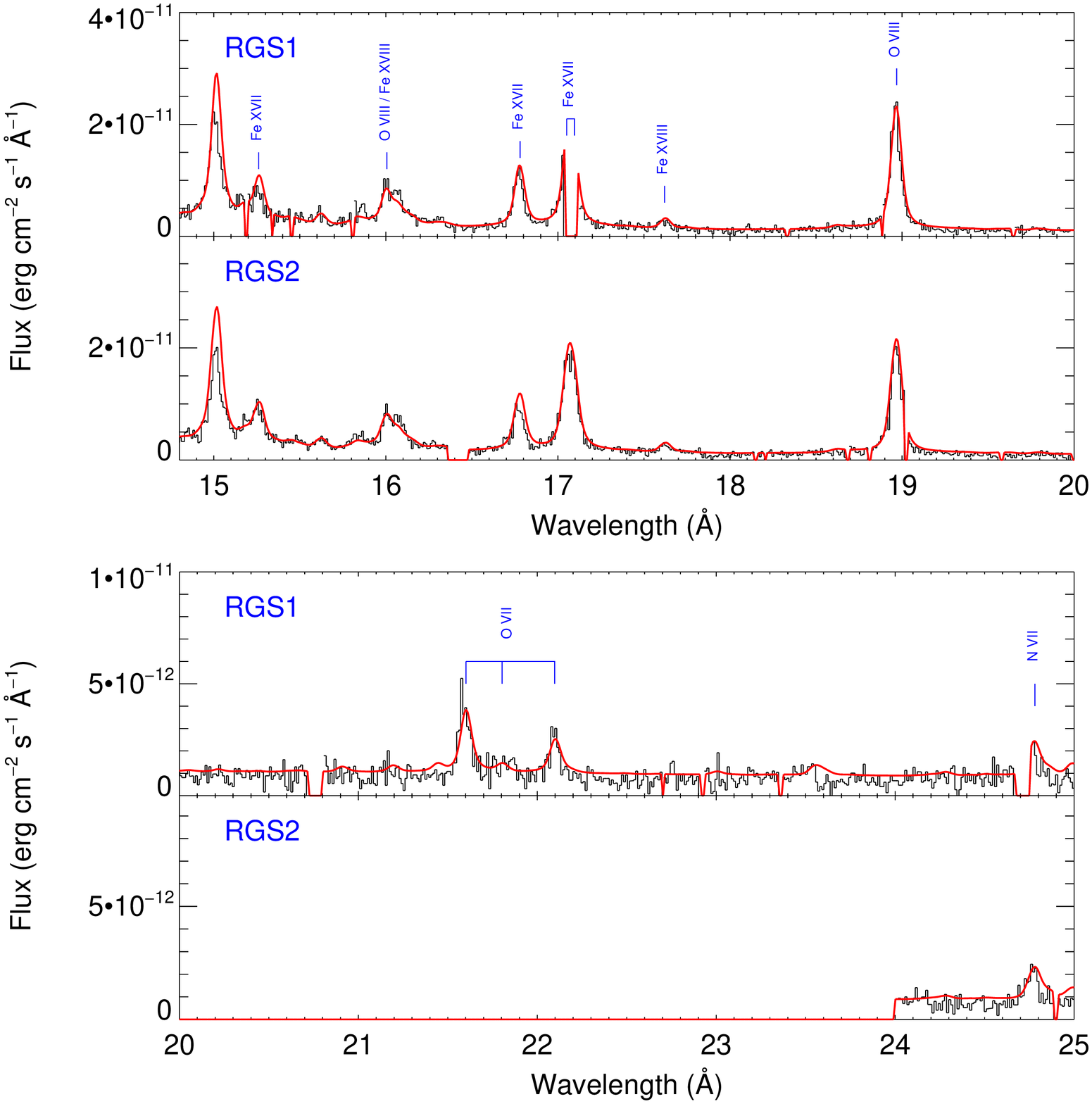}
\caption{\label{fig:xmmwchan2}}
\end{figure}
%\newpage

\clearpage
%%%%%%%%%%%%%%%%%%%%%%%%%%%%%%%%%%%%%%%%%%%%%%%%%%%%%%%%%%%%%%%%%%%%%%%%%%%%%%
\addtocounter{figure}{-1}
\addtocounter{subfigure}{1}
\begin{figure}
\centering
\includegraphics[width=\textwidth]{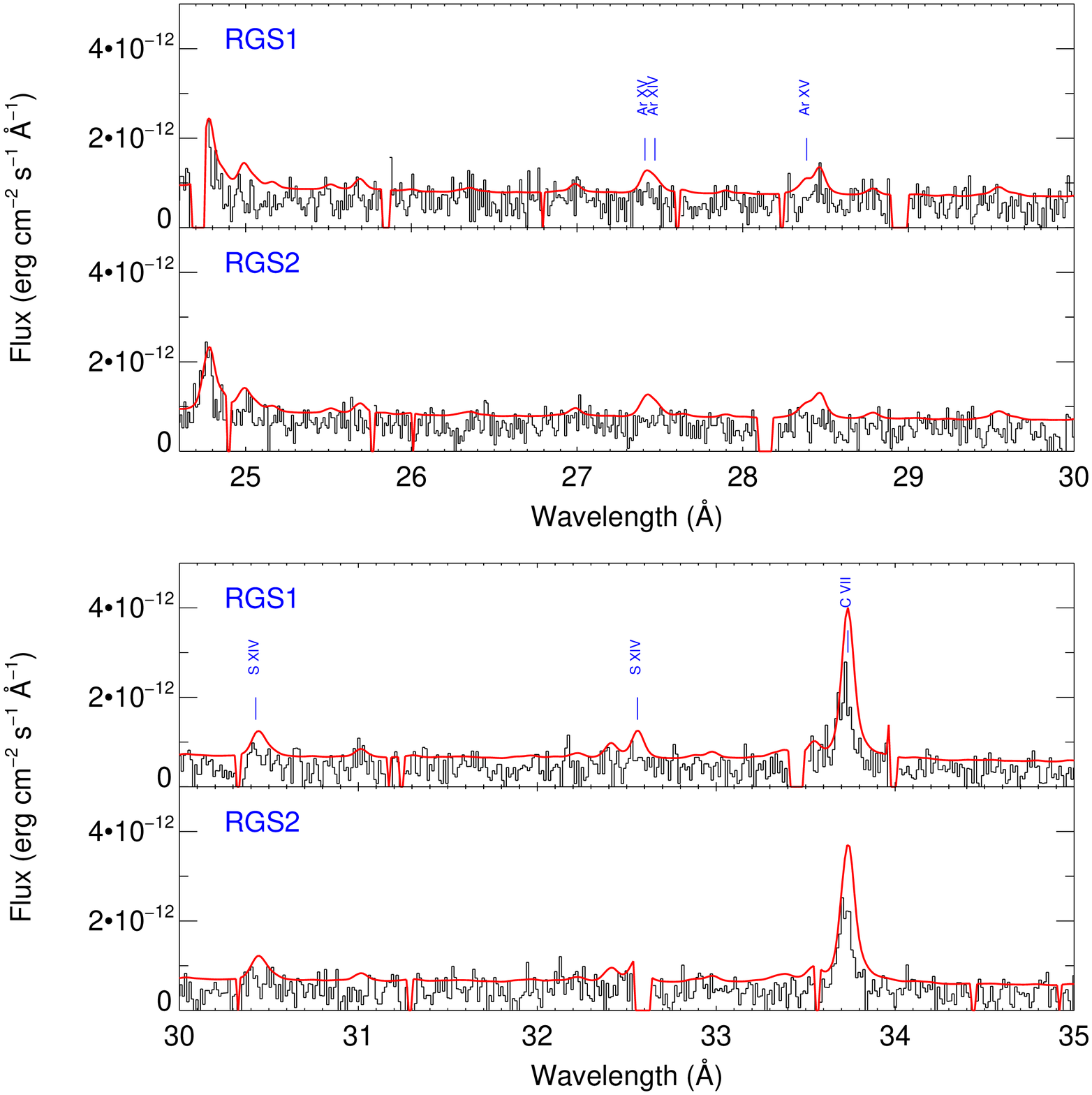}
\caption{\label{fig:xmmwchan3}}
\end{figure}
%\newpage

\clearpage

%%%%%%%%%%%%%%%%%%%%%%%%%%%%%%%%%%%%%%%%%%%%%%%%%%%%%%%%%%%%%%%%%%%%%%%%%%%%%%
%\renewcommand{\thefigure}{\arabic{figure}\alph{subfigure}}
\setcounter{subfigure}{1}
\begin{figure}
\centering
\includegraphics[width=\textwidth]{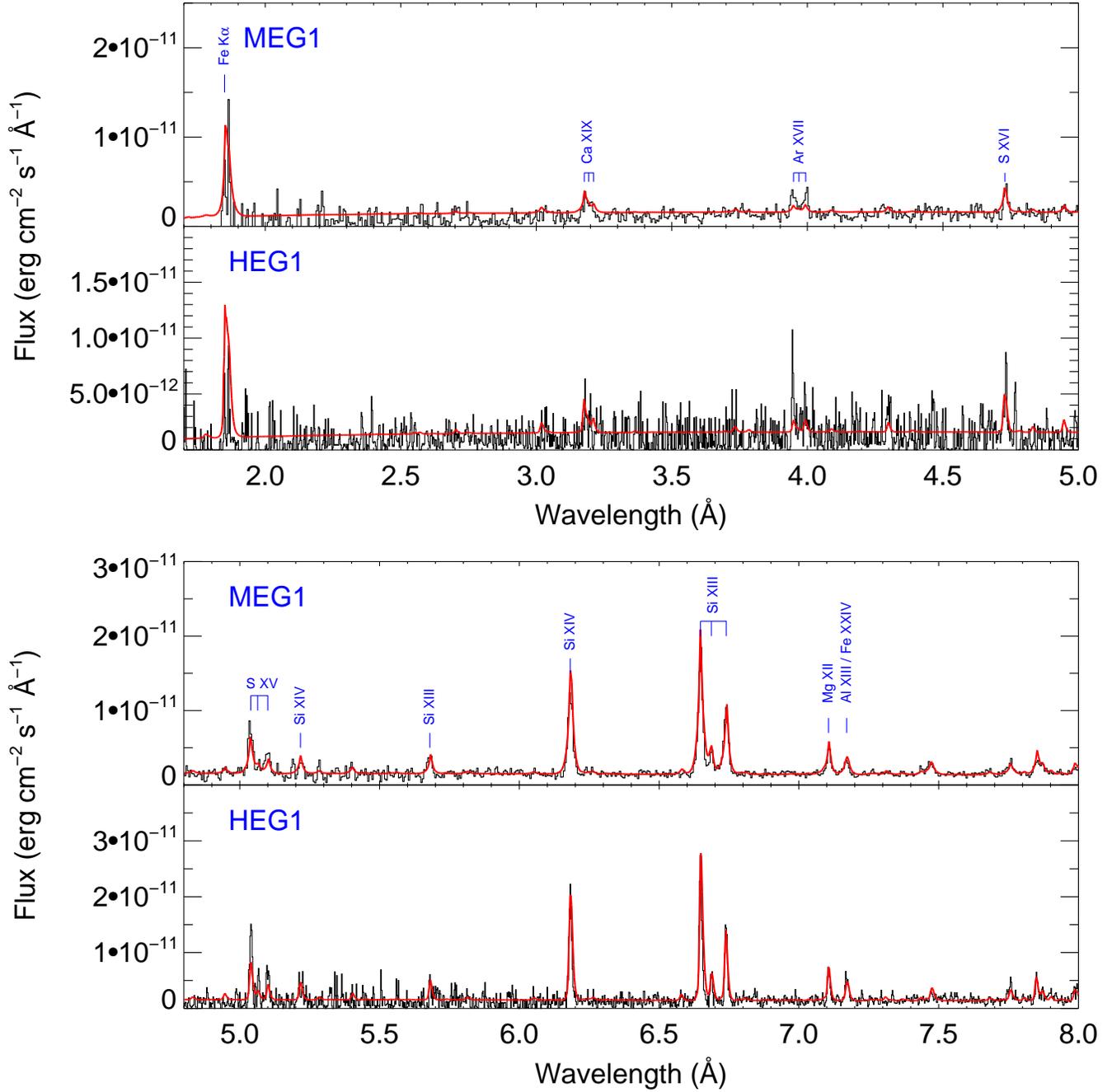}
\caption{($a$) \chan\  MEG1 and HEG1 fluxed spectra (uncorrected for interstellar absorption which is in any case negligible in
this wavelength range) compared with our \xmm\  EPIC MOS2+RGS (301) best-fit model 
(superposed line). ($b$) Same as
($a$), but for $8-14$~\AA. ($c$) Same as ($a$) but for $14-26$~\AA. The $17-26$~\AA\ 
panels show the MEG data only.\label{fig:chanwxmm1}}
\end{figure}
%\newpage

\clearpage

%%%%%%%%%%%%%%%%%%%%%%%%%%%%%%%%%%%%%%%%%%%%%%%%%%%%%%%%%%%%%%%%%%%%%%%%%%%%%%
\addtocounter{figure}{-1}
\addtocounter{subfigure}{1}
\begin{figure}
\centering
\includegraphics[width=\textwidth]{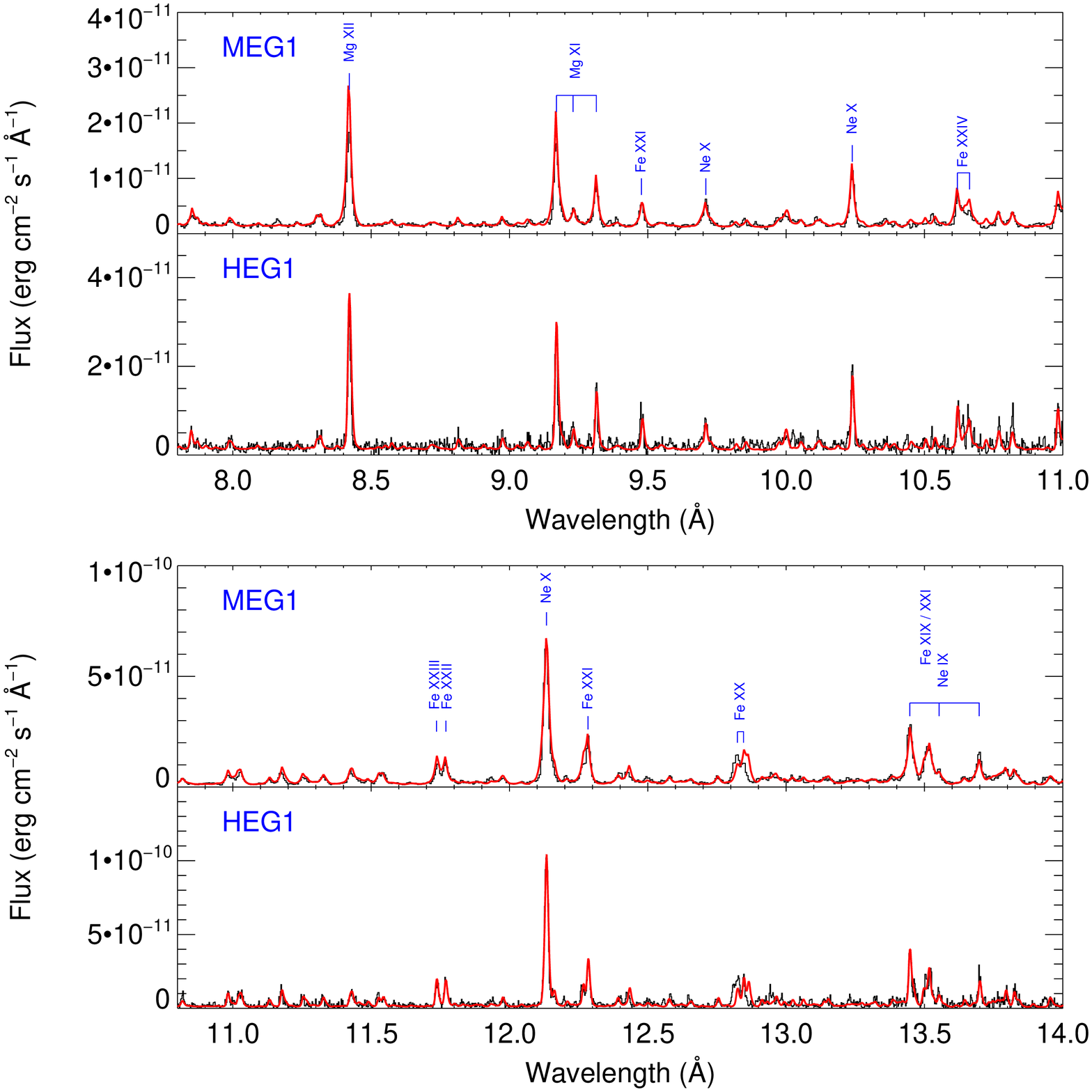}
\caption{\label{fig:chanwxmm2}}
\end{figure}
%\newpage

\clearpage

%%%%%%%%%%%%%%%%%%%%%%%%%%%%%%%%%%%%%%%%%%%%%%%%%%%%%%%%%%%%%%%%%%%%%%%%%%%%%%
\addtocounter{figure}{-1}
\addtocounter{subfigure}{1}
\begin{figure}
\centering
\includegraphics[width=\textwidth]{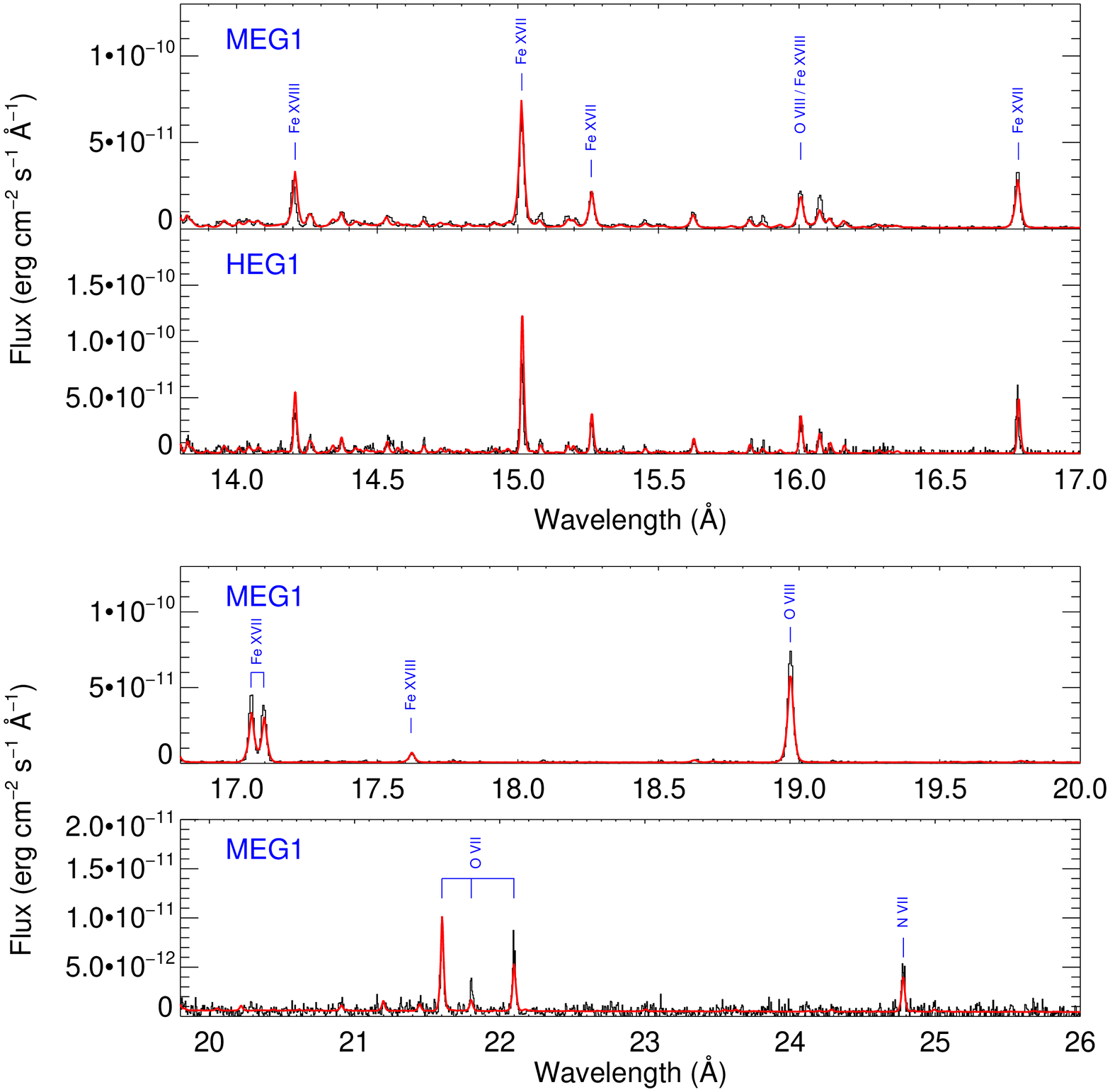}
\caption{\label{fig:chanwxmm3}}
\end{figure}
%\newpage

%%%%%%%%%%%%%%%%%%%%%%%%%%%%%%%%%%%%%%%%%%%%%%%%%%%%%%%%%%%%%%%%%%%%%%%%%%%%%%
\renewcommand{\thefigure}{\arabic{figure}}


\clearpage

\begin{thebibliography}{}
\bibitem[Agrawal et al.(1985)]{agrawal85} Agrawal, 
P.~C., Markert, T.~H., \& Riegler, G.~R.\ 1985, \mnras, 213, 761 

\bibitem[Agrawal et al.(1986)]{agrawal86} Agrawal, 
P.~C., Rao, A.~R., \& Riegler, G.~R.\ 1986, \mnras, 219, 777 

\bibitem[Agrawal et al.(1981)]{agrawal81} Agrawal, 
P.~C., White, N.~E., \& Riegler, G.~R.\ 1981, \mnras, 196, 73P 

\bibitem[Antunes et al.(1994)]{antunes94} Antunes, A., Nagase, F., \& White, N.~E.\ 1994, \apjl, 436, L83 

\bibitem[Arge et al.(1998)]{arge98}
Arge, C.~N., \& Mullan, D.~J. 1998 Sol. Phys, 182, 293

\bibitem[Arnaud(1996)]{arnaud96} Arnaud, K.~A. 1996, in ASP Conf. Ser. 101, Astronomical Data 
Analysis Software and Systems V, ed. G.~H. Jacoby \& J. Barnes (San Francisco:
ASP), 1

\bibitem[Audard et al.(2001b)]{audard01b} Audard, M., Behar, E., G\"udel, M.,
Raassen, A.~J.~J., Porquet, E., Mewe, R., Foley, C.~R., \& Bromage, G.~E. 2001b,
A\&A, 365, L329

\bibitem[Audard et al.(2001a)]{audard01a} Audard, 
M., G{\" u}del, M., \& Mewe, R.\ 2001a, \aap, 365, L318 

\bibitem[Audard et al.(2003)]{audard03} Audard, M., G\"udel, M., Sres, A.,
Raassen, A.~K.~K., \& Mewe, R. 2003, A\&A, 398, 1137

\bibitem[Audard et al.(2004)]{audard04} Audard, M., Telleschi, A., G\"udel, M., 
Skinner, S.~L., Pallavicini, R., \& Mitra-Kraev, U.  2004, \apj, 617, 531

\bibitem[Bhatia \& Doschek(1992)]{bhatia92} Bhatia, A.~K.~\& 
Doschek, G.~A.\ 1992, Atomic Data and Nuclear Data Tables, 52,


\bibitem[Bakos(1984)]{bakos84} Bakos, G.~A. 1984, AJ, 89, 1740


\bibitem[Barden(1985)]{barden85} Barden, S.~C. 1985, ApJ, 295, 162

\bibitem[Brinkman et al.(2001)]{brinkman01} Brinkman, A.~C., et al. 2001, 
A\&A, 365, L324

\bibitem[Brown et al.(2001)]{brown01} Brown, G.~V., Beiersdorfer, P., Chen, H., Chen, M.~H.,
\& Reed, K.~J. 2001, \apj, 557, L75

\bibitem[Brown et al.(1998)]{brown98} Brown, G.~V., Beiersdorfer, P., Liedahl,
D.~A., Wildmann, K., \& Kahn, S.~M. 1998, ApJ, 502, 1015

\bibitem[Chen \& Pradhan(2002)]{chen02} Chen, G.~X., \& Pradhan, A.~K.\ 2002, Physical Review Letters, 89, 013202 

\bibitem[Chen et al.(2003)]{chen03} Chen, G., Pradhan, A.~K., \& Eissner, W.\ 2003, Journal of Physics B Atomic Molecular Physics, 36, 453 

\bibitem[Craig \& Brown(1976a)]{craig76a} Craig, I.~J.~D.~\& 
Brown, J.~C.\ 1976a, \aap, 49, 239 

\bibitem[Craig \& Brown(1976b)]{craig76b} --------- 1976b, \nat, 264, 340 

\bibitem[den Herder et al.(2001)]{denherder01} den Herder, J.~W., et al. 2001, 
A\&A, 365, L7

\bibitem[Drake et al.(2001)]{drake01} Drake, J.~J., Brickhouse, 
N.~S., Kashyap, V., Laming, J.~M., Huenemoerder, D.~P., Smith, R., \& 
Wargelin, B.~J.\ 2001, \apjl, 548, L81 

\bibitem[Drake et al.(1995)]{drake95} Drake, J.~J., 
Laming, J.~M., \& Widing, K.~G.\ 1995, \apj, 443, 393 

\bibitem[Drake et al.(1997)]{drake97} ---------\ 1997, \apj, 478, 403 

\bibitem[Drake et al.(1989)]{drake89} Drake, S.~A., Simon, T., Linsky, K. 1989,
ApJS, 71, 905

\bibitem[Feldman(1992)]{feldman92} Feldman, U.\ 1992, Physica 
Scripta Volume T, 46, 202 

\bibitem[Fournier \& Hansen(2005)]{fournier05} Fournier, K.~B., \& Hansen, S.~B.\ 2005, \pra, 71, 012717 

\bibitem[Garcia-Alvarez et al.(2005)]{garcia05} Garcia-Alvarez, D., Drake, J.~J., Lin, L., Kashyap, V.~L., \& Ball, B. 2005, 621, 1009

\bibitem[Grevesse \& Sauval(1998)]{grev98} Grevesse, N., \& Sauval, A.~J. 1998,
Space Sci.~Rev., 85, 161

\bibitem[G{\" u}del et al.(2001a)]{guedel01a} G{\" u}del, M., et 
al.\ 2001a, \aap, 365, L336 

\bibitem[G{\" u}del et al.(2001b)]{guedel01b} G{\" u}del, M., 
Audard, M., Magee, H., Franciosini, E., Grosso, N., Cordova, F.~A., 
Pallavicini, R., \& Mewe, R.\ 2001b, \aap, 365, L344 


\bibitem[G{\" u}del et al.(2002)]{guedel02} G{\" u}del, M., 
Audard, M., Sres, A., Wehrli, R., Behar, E., Mewe, R., Raassen, A.~J.~J., 
\& Magee, H.~R.~M.\ 2002, ASP Conf.~Ser.~277: Stellar Coronae in the 
\chan\   and \xmm\  Era, 497 

\bibitem[H\'enoux(1995)]{henoux95} 
H\'enoux, J.-C. 1995, Adv.~Space~Res., 15, 23

\bibitem[H\'enoux(1998)]{henoux98} 
H\'enoux, J.-C. 1998, Space~Sci.~Rev., 85, 215

\bibitem[Huenemoerder et al.(2003)]{huenemoerder03} Huenemoerder, D.~P., Canizares, 
C.~R., Drake, J.~J., \& Sanz-Forcada, J.\ 2003, \apj, 595, 1131 

\bibitem[Huenemoerder et al.(2001)]{huenemoerder01} 
Huenemoerder, D.~P., Canizares, C.~R., \& Schulz, N.~S.\ 2001, \apj, 559, 
1135 

\bibitem[Jansen et al.(2001)]{jansen01} Jansen, F., et al. 2001, A\&A, 365, L1

\bibitem[Laming(2004)]{laming04} Laming, J.~M.  ApJ, 614, 1063

\bibitem[Laming \& Drake(1999)]{laming99} Laming, J.~M.~\& 
Drake, J.~J.\ 1999, \apj, 516, 324 

\bibitem[Laming et al.(1995)Laming, Drake, \& Widing]{laming95} Laming, 
J.~M., Drake, J.~J., \& Widing, K.~G.\ 1995, \apj, 443, 416 

\bibitem[Laming et al.(2000)]{laming00} Laming, J.~M., Kink, I., Takacs, E., et
al. 2000, ApJ, 545, L161

\bibitem[Lee et al.(2001)]{lee01} Lee, J.~C., Ogle, P.~M., Canizares, C.~R., Marshall, H.~L., Schulz, N.~S., 
Morales, R., Fabian, A.~C., \& Iwasawa, K.\ 2001, \apjl, 554, L13

\bibitem[Lepson et al.(2003)]{lepson03} Lepson, J.~K., Beiersdorfer, P., Behar, E., 
\& Kahn, S.~M.\ 2003, \apj, 590, 604 

\bibitem[Lumb et al.(2000)]{lumb00} Lumb, D. H., Gondoin, Ph., \& Turner, M. J.
 L. 2000, SPIE, 4140, L22

\bibitem[McKenzie(2000)]{mckenzie00}
 McKenzie, J.~F.\ 2000, Sol. Phys., 196, 329 

\bibitem[Meyer(1985)]{meyer85} Meyer, J.~P. 1985, ApJS, 57, 173

\bibitem[Morrison \& McCammon(1983)]{morrison83} Morrison, R., \& McCammon, D.
1983, ApJ, 270, 119

\bibitem[Ness et al.(2003)]{ness03} Ness, J.-U., Schmitt, 
J.~H.~M.~M., Audard, M., G{\" u}del, M., \& Mewe, R.\ 2003, \aap, 407, 347 

\bibitem[Osten et al.(2000)]{osten00} Osten, R.~A., Brown, A., Ayres, T.~R.,
Linsky, J.~L., Drave, S.~A., Gagn\'e, M., \& Stern, R.~A. 2000, ApJ, 544, 953

\bibitem[Osten et al.(2003)]{osten03} Osten, R.~A., Ayres, T.~R., Brown, A.,
Linsky, J., \& Krishnamurthi, A. 2003, ApJ, 582, 1073

\bibitem[Pasquini et al.(1989)]{pasquini89} 
Pasquini, L., Schmitt, J.~H.~M.~M., \& Pallavicini, R.\ 1989, \aap, 226, 
225 

\bibitem[Perryman et al.(1997)]{perryman97} Perryman, M.~A.~C., et al. 1997, \aap, 33, L49


\bibitem[Porquet et al.(2001)]{porquet01} Porquet, D., Mewe, R., Dubau, J.,
Raassen, A.~J.~J., Kaastra, J.~S. 2001, A\&A, 376, 1113

\bibitem[Raassen et al.(2002)]{raassen02} Raassen, A.~J.~J., et 
al.\ 2002, \aap, 389, 228 

\bibitem[Raassen et al.(2003)]{raassen03} 
Raassen, A.~J.~J., Mewe, R., Audard, M., \& G{\" u}del, M.\ 2003, \aap, 
411, 509 

\bibitem[Sako et al.(2003)]{sako03} Sako, M., et al.\ 2003, \apj, 596, 114 

\bibitem[Sanz-Forcada et al.(2003b)]{sanz03b} 
Sanz-Forcada, J., Brickhouse, N.~S., \& Dupree, A.~K.\ 2003b, \apjs, 145, 
147 

\bibitem[Sanz-Forcada et al.(2004)]{sanz04} 
Sanz-Forcada, J., Favata, F., \& Micela, G.\ 2004, \aap, 416, 281 

\bibitem[Sanz-Forcada et al.(2003a)]{sanz03a} 
Sanz-Forcada, J., Maggio, A., \& Micela, G.\ 2003a, \aap, 408, 1087 

\bibitem[Schmitt et al.(1996)]{schmitt96} 
Schmitt, J.~H.~M.~M., Stern, R.~A., Drake, J.~J., \& Kuerster, M.\ 1996, 
\apj, 464, 898 

\bibitem[Schwadron et al.(1999)]{schwadron99}
Schwadron, N.~A., Fisk, L.~A., \& Zurbuchen, T.~H. 1999, ApJ, 521, 859

\bibitem[Smith et al.(2001)]{smith01} Smith, R.~K., Brickhouse, N.~S., Liedahl,
 D.~A., \& Raymond, J.~C. 2001, ApJ, 556, L91
 
\bibitem[Stern et al.(1992)]{stern92} Stern, R.~A., Uchida, Y., 
Walter, F., Vilhu, O., Hannikainen, D., Brown, A., Veale, A., \& Haisch, 
B.~M.\ 1992, \apj, 391, 760 

\bibitem[Strassmeier et al.(1993)]{strassmeier93} Strassmeier, K.~G., Hall, D.~S.,
Fekel, F.~C., \& Scheck, M. 1993, A\&AS, 100, 173

\bibitem[Strassmeier \& Rice(2003)]{strassmeier03} Strassmeier, K.~G., Rice, J.~B.
2003, A\&A, 399, 315

\bibitem[Telleschi et al.(2005)]{telleschi05} Telleschi, A., G\"udel, M., Briggs, K.~R., Audard, M., 
 Ness, J.-U., \& Skinner, S.~L. 2004, \apj, 622, 653
 
\bibitem[Testa et al.(2004)]{testa04} Testa, P., Drake, J.~J., Peres, G. 2004, \apj, 617, 508 
 
\bibitem[Turner et al.(2001)]{turner01} Turner, M.~J.~L., et al. 2001, A\&A, 
365, L27

\bibitem[van den Besselaar et al.(2003)]{vandenbesselaar03} van den 
Besselaar, E.~J.~M., Raassen, A.~J.~J., Mewe, R., van der Meer, R.~L.~J., 
G{\" u}del, M., \& Audard, M.\ 2003, \aap, 411, 587 

\bibitem[van den Oord et al.(1988)]{vandenoord88}
den Oord, G.~H.~J., Mewe, R., \& Brinkman, A.~C.\ 1988, \aap, 205, 181 

\bibitem[White et al.(1994)]{white94} White, N.~E., et al.\ 1994, \pasj, 46, L97 
\end{thebibliography}
\end{document}